# Time evolution of the process of doping of solids by plasma-ion beams

Andrzej Horodeński, Cezary Pochrybniak, Roch Kwiatkowski

*National Centre for Nuclear Research, Świerk, Poland*

***Abstract.*** *Irradiation of a solid with intense plasma-ion beams produced by Rod Plasma Injector is a strongly nonequilibrium process, which enables achieving a number of effects which are impossible to be achieved with other methods – improvement of ceramics wettability, fabrication of stable copper-ceramics interfaces and stable Ni-Cu and Al-Cu interfaces, improvement of tribological properties and high temperature oxidation resistance of stainless steel, photovoltaic junction formation, and many others. In the paper, the process of plasma-ion beam propagation regarding its time and energy distributions and the process of ion penetration of solids, resulting with ion implementation and temperature growth have been analyzed mathematically on basis of experimental data. Results of numerical calculations have been presented concerning temperature and dopant density time evolution.*

***Keywords:*** *plasma, ion beam, implantation, doping, time evolution, energy distribution, stopping power*

## Introduction

Irradiation of solids with so-called plasma-ion beams is a strongly nonequilibrium process, which enables a number of effects that are impossible to achieve with other methods [1]. These include improvement of wettability of ceramics [2,4,8], production of stable copper-ceramics interfaces [3], production of stable Ni-Cu and Al-Cu interfaces [10], improvement of properties of zirconium alloys [13], improvement of tribological properties of stainless steel [5], improvement of high temperature oxidation resistance of stainless steel [6], modification of superconducting and electrical properties of Mg–B structures [7], improvement of manganese distribution in Si with $He^+$ and $H^+$ plasma pulse irradiation [9], doping metals with nitrogen [11], and photovoltaic junction formation [12].

The concept of the so-called Rod Plasma Injector (RPI) was proposed by Michał Gryziński [14,15] and verified experimentally [16-19], including the recently set in operation IBIS-II RPI-type plasma generator [20]. The RPI is a very effective source of intense plasma-ion beams emitted in the form of short pulses of a duration of few microseconds, plasma density of the order of $10^{16}$ cm$^{-3}$, and ion energy distribution up to several hundreds of keV – see Figs. 1-2. The process of ion propagation was initially analyzed in lit. [22].

# 1. Ion-plasma beam propagation

A complete set of ion emitting source parameters must consist of the following data:

1. total number of ions,
2. ion energy distribution,
3. ion start time distribution.

For numerical calculations (whose results are presented at the final part of the article), the following input data have been used:

1. energy distribution $f_e(\varepsilon)$ has the form shown in Fig. 1,
2. ion start time distribution $\Phi(\tau)$ has the form shown in Fig 2,
3. having been accelerated within the plasma generator, ions move straightforward with uniform motion,
4. number of ions $N_0 = 5*10^{12}$ cm$^{-2}$,
5. target: silicon *Si*.

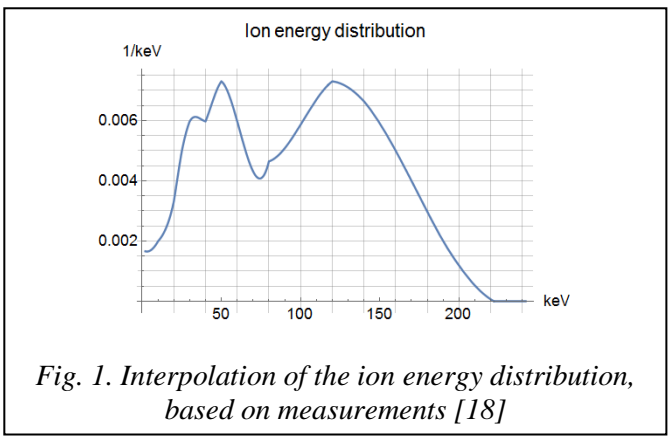

*Fig. 1. Interpolation of the ion energy distribution, based on measurements [18]*

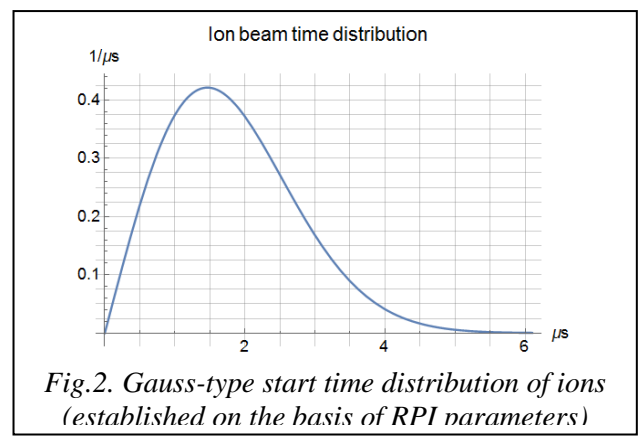

*Fig.2. Gauss-type start time distribution of ions (established on the basis of RPI parameters)*

With initial conditions listed above, the total on-target energy density is about 0.08 J/cm$^2$.

An ion source can be understood as some surface $\Sigma$, which emits ions in accordance with the energy distribution $f_e(\varepsilon)$, and time distribution $\Phi(\tau)$.

To begin with, we will consider an elementary ion "microsource" stretched over a surface element $d\Sigma$, emitting ions within a time interval $[\tau,\tau+d\tau]$ - see Fig. 3.

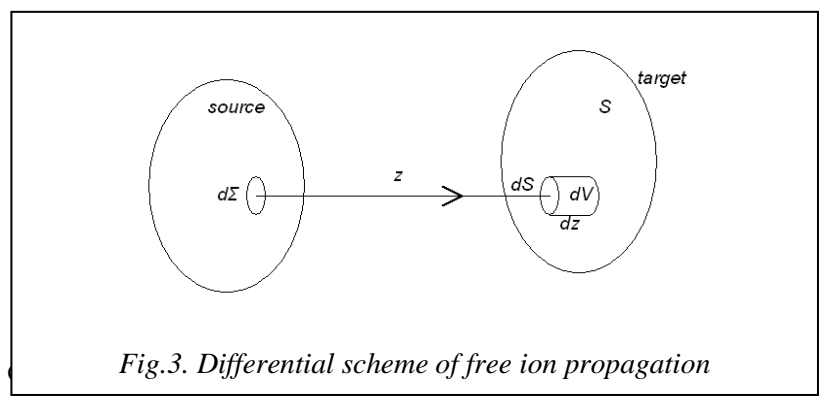

*Fig.3. Differential scheme of free ion propagation*

The number of ions

$$\delta N = N_0 \, d\Sigma \, \Phi(\tau) \, d\tau \tag{1}$$

The number of ions emitted from the "microsource" with an energy within [$\varepsilon, \varepsilon+d\varepsilon$] is:

$$dN = \delta N \, f_e(\epsilon) \, d\epsilon \tag{2}$$

Ions move from the source to a target along trajectories which can be expressed in terms of kinetic energy $\varepsilon$, flight time $\tau$, and position vector z:

$$\epsilon = \frac{mz^2}{2(t-\tau)^2} \tag{3}$$

In order to obtain expressions for on-target observables, one must transform the energy function $f_e$ into the respective time-of-flight distribution function at a fixed source-target distance:

$$f_\tau(t-\tau, z) = f_e\left(\frac{mz^2}{2(t-\tau)^2}\right) \frac{d\epsilon}{d(t-\tau)} = \frac{mz^2}{(t-\tau)^3} f_e\left(\frac{mz^2}{2(t-\tau)^2}\right) \quad (z = \text{const.}) \tag{4}$$

Now, the number of particles (from the "microsource") filling the volume element $dV$ within the time interval [$\tau, \tau+d\tau$] can be calculated as:

$$dN = N_0 \, d\Sigma \, f_\tau(t-\tau, z) \, d(t-\tau) \, \Phi(\tau) d\tau \tag{5}$$

Taking into account geometrical relations:

$$\frac{dz}{d(t-\tau)} = \frac{z}{t-\tau} = v, \quad dV = dz \, dS \tag{6}$$

we finally obtain the integral equation which connects ion source parameters with the on-target beam density:

$$n = \frac{dN}{dV} \rightarrow n(t,z) = \frac{dN}{dV} = m \, N_0 \int_0^t \frac{z}{(t-\tau)^2} f_e\left(\frac{mz^2}{2(t-\tau)^2}\right) \Phi(\tau) d\tau \tag{7}$$

The same can be done with flux and power:

$$dJ = dn \, v = dn \, \frac{z}{t-\tau} \rightarrow J(t,z) = mN_0 d\Sigma \int_0^t \frac{z^2}{(t-\tau)^3} f_e\left(\frac{mz^2}{2(t-\tau)^2}\right) \Phi(\tau) d\tau \tag{8}$$

$$dP = \frac{dn \, dV}{d(t-\tau) \, dS} \frac{mz^2}{2(t-\tau)^2} \rightarrow P(t,z) = \frac{1}{2} m^2 N_0 d\Sigma \int_0^t \frac{z^4}{(t-\tau)^5} f_e\left(\frac{mz^2}{2(t-\tau)^2}\right) \Phi(\tau) d\tau \tag{9}$$

The density of the energy deposited on a target is:

$$E(z) = \int_0^t P(t,z) dt \tag{10}$$

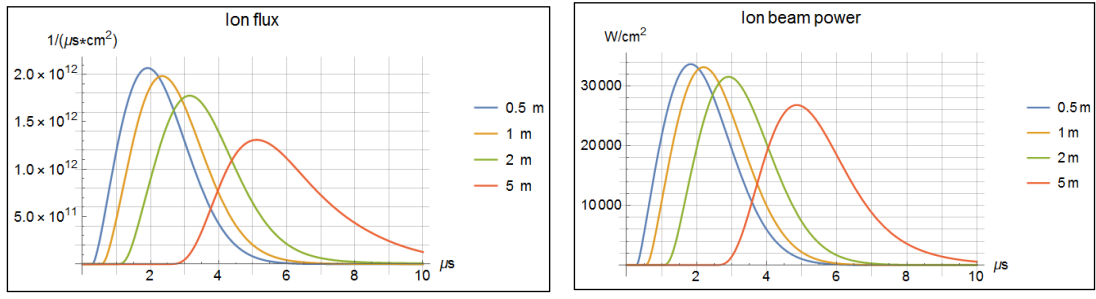

*Fig.4. Temporary shapes of the flux (left) and power (right) of the plasma-ion beam at different distances from the plasma source.*

## 2. Time evolution of doping process

The formulae derived above enable a numerical model to be constructed, which enables an immediate evaluation of the on-target ion flux shape at hand. Henceforth, it is possible to compute an evaluation of the implantation process parameters .

To achieve this goal, one has to "filter" the ion beam flux formula (8) using the energy dependent Dirac delta function:

$$j(\epsilon, t, z) = mN_0 \int_{-\infty}^{+\infty} J(t, z)\, \delta(\epsilon - \frac{mz^2}{2(t-\tau)^2}) d\epsilon \qquad (11)$$

From (11) we obtain, after some development, the time dependent energy distributed flux of ions approaching the target:

$$j(\epsilon, t, z) = mN_0 \Phi\left(t - \sqrt{\frac{mz^2}{2\epsilon}}\right) f_e(\epsilon) \qquad (12)$$

As seen above, the flux is elicited directly from ion source parameters, i.e. energy distribution, time distribution of the emission, and number of particles. An example of the flux of target approaching ions, computed at a series of time instants, is shown in Fig. 5.

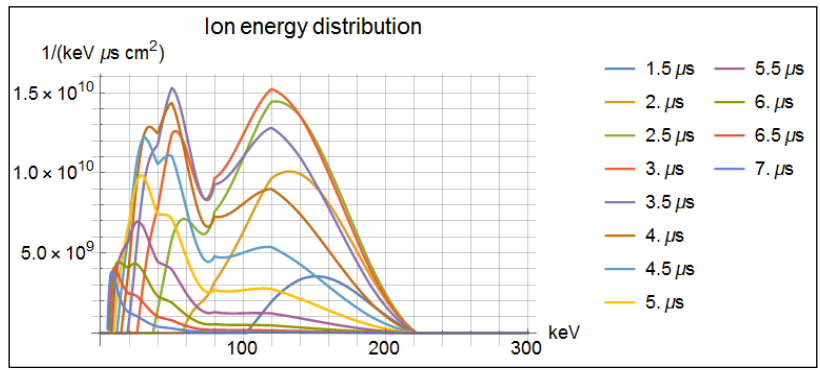

*Fig.5. History of energy distribution of the flux of $B^+$ ion beam approaching the target surface.*

Coupled with the parameters of the ion beam impinging the solid, formula (12) allows us to precisely compute details of the process of ion penetration of a target. As the result, we can achieve time dependent in-depth distributions of densities of doping ions, as well as a full history of the temperature distribution resulting from the energy released by ion irradiation.

During the solid penetration process, the temporary position of the ion depends on the material parameters and initial energy. It can be calculated with stopping power data [22]:

$$z(\epsilon_0, \epsilon) = -\int_{\epsilon_0}^{\epsilon} \frac{d\epsilon}{S(\epsilon)} \qquad (14)$$

where: $\epsilon_0$ – initial energy value,
$\epsilon$ – local energy value,
$S$ – stopping power function.

From (14) we immediately get the ion doping depth:

$$z_{dop}(\epsilon_0) = z(\epsilon_0, 0) \qquad (15)$$

Examples of a stopping power function shape of diverse ions penetrating a silicon wafer are shown in Fig. 6. An adequate dependence of the doping depth on the initial ion energy is shown in Fig. 7.

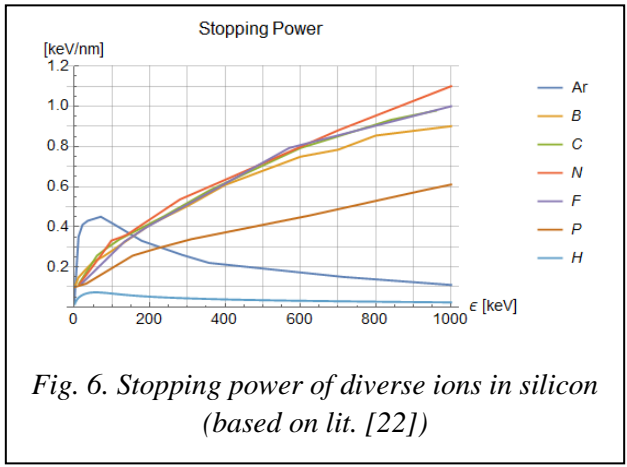

Fig. 6. Stopping power of diverse ions in silicon (based on lit. [22])

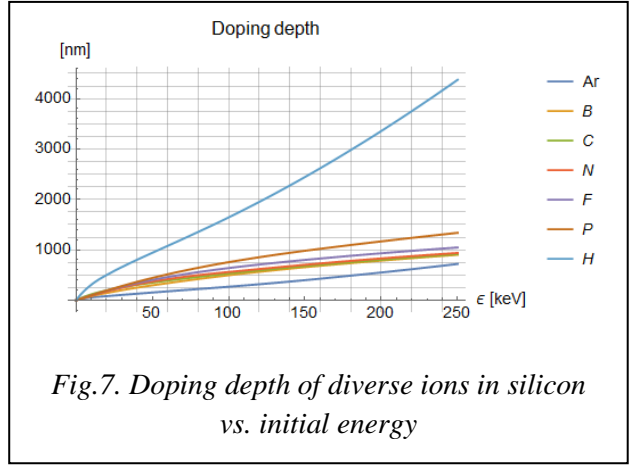

Fig.7. Doping depth of diverse ions in silicon vs. initial energy

The above data makes it possible to estimate the duration of a single ion doping process:

$$\tau = 2\,\delta\,(m/\varepsilon_0)^{1/2} \approx 10^{-12}\ s \qquad (16)$$

$\delta$ – penetration depth,
$\varepsilon_o$ – initial energy,
$m$ – ion mass.

As seen above, the duration of the doping process is several orders of magnitude less than half of an ion pulse. Therefore, it can be assumed that the process of penetration, i.e. from impacting to stopping, is in fact an immediate one, „out of time", within the actual time framework (on the order of 5 μs). This fact enables us to directly transform the initial energy of ions into an in-depth distribution. This can be done by an exchange of energy coordinate into position coordinate (prior to this, the reversed form of the function (15) must be found):

$$j(\epsilon_0, t) = j(\epsilon_0(z), t)\frac{d\epsilon_0(z_{dop})}{dz_{dop}} \qquad (17)$$

From this, we can compute the in-depth dopant concentration:

$$n(z,t) = \frac{dN}{dz_{dop}} = \int_0^t j(\epsilon_0(z_{dop}), \tau) \frac{d\epsilon_0(z_{dop})}{dz_{dop}} d\tau \qquad (18)$$

The calculated time evolution of a doping process is shown in Fig. 8.

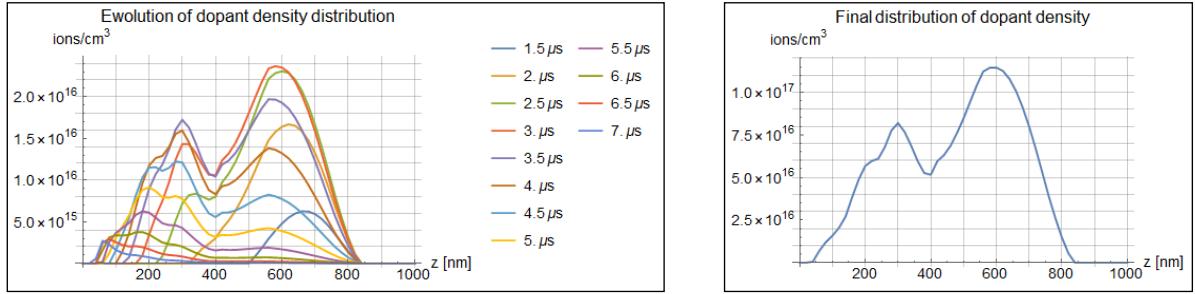

*Fig. 8. Left: time evolution of dopant density distribution. Right: final dopant density distribution ($B^+$ ions in silicon). Distance between plasma source and the target is 2 m.*

## 3. Time evolution of temperature distribution

As shown by formula (16), the ion transfer from the surface to a certain position is much faster than the doping process time scale, therefore the position dependent ion energy loss depends solely on the initial energy of the ion. Therefore, the energy deposited at the 'z' position is the sum of energies of all ions approaching this position within the time interval [$t,t+dt$].

Hence, the total energy transferred to a solid by a group of ions with the energy of [$\varepsilon,\varepsilon+d\varepsilon$] and time interval [$t,t+dt$] at the position 'z' is equal to (the function $\varepsilon(z, \varepsilon_0)$ is elicited from formula (14)):

$$dQ(z,t,\varepsilon_0) = dz\, dt\ j(\varepsilon_0, t, d) \frac{d\varepsilon(z,\varepsilon_0)}{dz} d\varepsilon_0 \qquad (19)$$

From (19) we immediately achieve the temperature time evolution:

$$T_{ref}(z,t) = \frac{dQ}{\delta\, \gamma\, dz} = \frac{1}{\delta\, \gamma} \int_0^t dt \int_{\varepsilon_0}^{\varepsilon} j(\varepsilon_0, t, d) \frac{d\varepsilon(z,\varepsilon_0)}{dz} d\varepsilon_0 \qquad (20)$$

*d – distance between source and target,*
*δ – target stuff density,*
*γ – specific heat.*

Expression (20) has an essential fault – it does not consider that the target substance melts. Consequently, it must be understood as only a sort of "reference temperature", which gives the basis for the final procedure of the real temperature computation which contains the process of melting. Finally, calculation is as follows:

$$T(z,t) = \begin{cases} T_{ref}(z,t) & \text{if } T_{ref} < T_{melt} \\ T_{melt} & \text{if } T_{melt} < T_{ref} < T_{melt} + \sigma_{melt}/\gamma \\ T_{ref}(z,t) - \sigma_{melt}/\gamma & \text{if } T_{ref} > T_{melt} + \sigma_{melt}/\gamma \end{cases} \quad (21)$$

$T_{melt}$ – melting point,
$\sigma_{melt}$ – melting heat,
$\gamma$ – specific heat [23].

An example of calculation of in-depth temperature distribution is shown in Fig. 9.

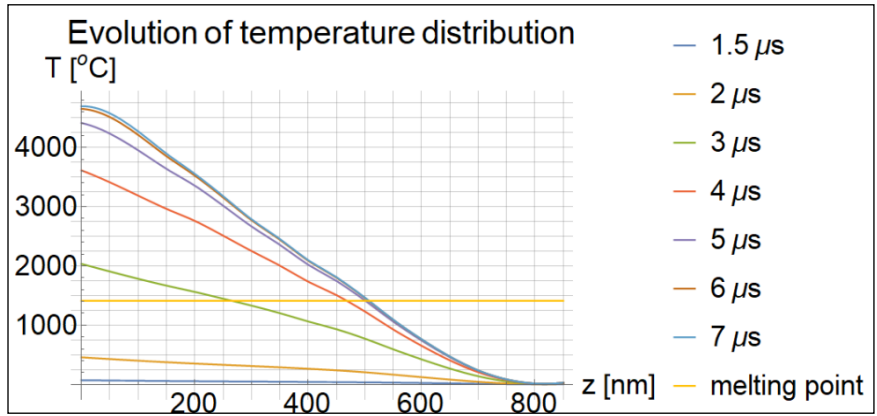

Fig. 9. Time evolution of temperature distribution ($B^+$ in silicon).
Distance between plasma source and the target is 2 m.

Final results of silicon wafer irradiation with plasma-ion beams consisting of diverse elements are shown in Figs. 10 and 11. As seen, parameters of doping processes with argon, boron, nitrogen, carbon, or fluorine ions are quite similar. At the same time, the hydrogen plasma reacts to silicon in quite a different way.

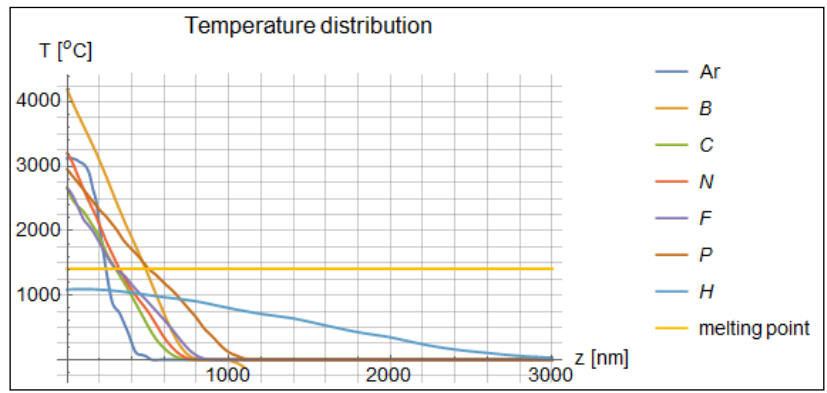

Fig. 10. Final temperature distributions of silicon target irradiated with different ion beams.

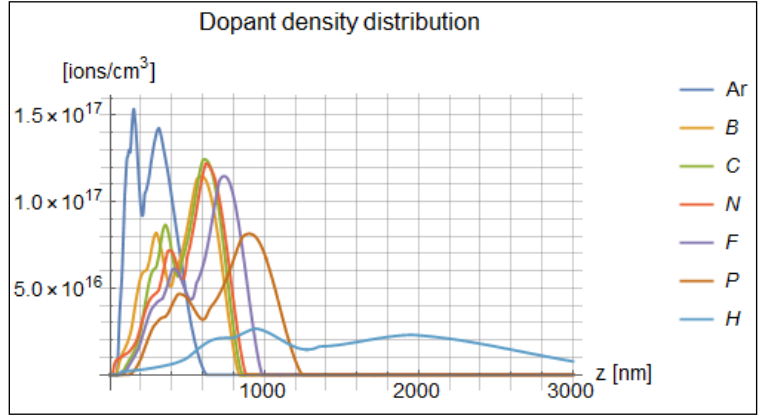

*Fig. 11. Final dopant density distributions at silicon target irradiated with different ion beams.*

## 4. Energy dissipation

Accuracy of the presented model is disturbed by at least five dissipative processes: evaporation, infrared emission, heat flow, deionization and sputtering.

i. Evaporation

$$\frac{\partial q}{\partial t} \approx p\sqrt{(3kT/m)} \approx 1.8*10^{-3} \ J/(\mu s \ cm^2)$$

*p – equilibrium vapour pressure (Si: $10^4$ Pa at 2700 °C) [25]*
*m – atom mass.*

ii. Infrared emission

$$\frac{\partial q}{\partial t} = \varepsilon \sigma T^4 \approx 2.3*10^{-4} \ J/(\mu s \ cm^2) \ (at \ T=2700 \ °C)$$

*ε – emissivity (Si: ε ≈ 0,5 [24])*
*σ – Stefan-Boltzmann constant.*

iii. Heat flow:

$$\frac{\partial q}{\partial t} = -\lambda \frac{\partial T}{\partial z} \qquad (22)$$

*T – temperature,*
*q – heat flux,*
*λ – heat conductivity (Si: 148 W/(mK).*

The heat flux, evaluated on the basis of the heat flow equation (22) at the temperature gradient of ca. 200 K/µm (taken from Fig. 9), is:

$$\frac{\partial q}{\partial t} \approx 3*10^{-5} \ J/(\mu s \ cm^2)$$

iv. Deionization

Penetrating a solid body, ions are submitted to deionization process. Therefore they release the energy less or equal to their ionization potential:

$$\frac{\partial q}{\partial t} \approx \varepsilon_i\, N_0/\Delta t \approx 5.8*10^{-7}\ J/(\mu s\ cm^2)$$

*$\varepsilon_i$ – ionization potential of boron: $\varepsilon=8,3\ eV$.*

v. Sputtering

$$N_{sputt} = \varepsilon_s*N_0 \approx 1.6*10^{12}\ cm^{-2} \qquad (<1\ atomic\ layer)$$

*$\varepsilon_s$ – sputter yield, ($\varepsilon_s \approx 0.33$, B:Si, 1-10 keV)* [26]

Within the analysed set of parameters sputtering process affects the solid surface within the extent of only 1 atomic layer, and therefore it can be regarded insignificant.

Taking into account that the incoming energy flux is on the order of $2*10^{-2}\ J/(\mu s\ cm^2)$, we may assume that the dissipative processes mentioned above are too slow to have a significant influence on the process of ion doping. However, above 2700 °C the influence of the evaporation heat flow (with a silicon target) is becoming indispensable.

## 5. Conclusions

Having the set of computation tools defined above, and within the assumed framework of initial parameters including energy distribution, emission pulse duration and pulse shape, number of ions, and the distance between plasma source and irradiated target, one can perform a credible comparison of doping processes of different ions. Further analysis will concern a significantly enlarged temperature (above 2500 °C) and time frames (from microseconds to milliseconds and more) in which the dissipative processes as well as dopant migration in the liquid phase of a solid will be taken into account.

# References

[1] J. Piekoszewski, Z. Werner, C. Pochrybniak, J. Langner, M. Gryzinski, A. Horodenski, *Pulse Ion Beam Doping and Modification of Solids*, physica status solidi (a) **112**(2):757-760, 1989

[2] M. Barlak, J. Piekoszewski, J. Stanisławski, Z. Werner, K. Borkowska, *The effect of intense plasma pulse pre-treatment on wettability in ceramic–copper system*, Fusion Engineering and Design 82 (15-24), 2524-2530

[3] M. Barlak, W. Olesinska, J. Piekoszewski, Z. Werner, M. Chmielewski, *Ion beam modification of ceramic component prior to formation of AlN-Cu joints by direct bonding process,* Surface and Coatings Technology 201 (19-20), 8317-8321

[4] M. Barlak, J. Piekoszewski, Z. Werner, J. Stanislawski, *Wettability improvement of carbon ceramic materials by mono and multi energy plasma pulses,* Surface and Coatings Technology 203 (17-18), 2536-2540

[5] B. Sartowska, J. Piekoszewski, L. Waliś, J. Senatorski, M Barlak, W Starosta, *Improvement of tribological properties of stainless steel by alloying its surface layer with rare earth elements using high intensity pulsed plasma beams*, Surface and Coatings Technology 205, S124-S127

[6] J. Piekoszewski, B. Sartowska, M. Barlak, P. Konarski, L. Dąbrowski, *Improvement of high temperature oxidation resistance of AISI 316L stainless steel by incorporation of Ce–La elements using intense pulsed plasma beams,* Surface and Coatings Technology **206** (5), 854-858

[7] J. Piekoszewski, W. Kempiński, M. Barlak, J. Kaszyński, J. Stanisławski, *Superconducting and electrical properties of Mg–B structures formed by implantation of magnesium ions into the bulk boron followed by pulse plasma treatment,* Vacuum **81** (10), 1398-1402

[8] M. Barlak, J. Piekoszewski, Z. Werner, B. Sartowska, Wojciech Starosta, J. Kierzek, C. Pochrybniak, E. Kowalska, *Wettability of carbon and silicon carbide ceramics induced by their surface alloying with Zr and Cu elements using high intensity pulsed plasma beams,* Nukleonika **57**(4),477-483, 2012

[9] Z. Werner, C. Pochrybniak, M. Barlak, J. Piekoszewski, A. Korman, R. Heller, W. Szymczyk, K. Bochenska, *Implanted manganese redistribution in Si after He+ irradiation and hydrogen pulse plasma treatment,* Nukleonika,- **56** (1), 5-8, 2011

[10] J. Langner, J. Piekoszewski, C. Pochrybniak, F. Rosatelli, S. Rizzo, J. Kucinski, A. Miotello, Lina Alejandra Guzmán, Paolo Lazzeri, *Deposition by pulsed erosion of nickel and aluminum on copper,* Surface and Coatings Technology **66** (1-3), 300-304, 1994

[11] J. Piekoszewski, J. Langner, J. Białoskórski, B. Kozłowska, C. Pochrybniak, Z. Werner, M. Kopcewicz, L. Waliś, A. Ciurapiński, *Introduction of nitrogen into metals by high intensity pulse ion beams*, Nuclear Instruments and Methods in Physics, Research Section B Beam Interactions with Materials and Atoms **80–81**:344–347, 1993

[12] J. Piekoszewski, M. Gryzinski, J. Langner, Z. Werner, G.C. Huth, *A new approach to photovoltaic junction formation by using pulse implantation doping technique,* J. Phys. France **43**, 1353-1358 (1982)

[13] B. Sartowska, W. Starosta, M. Barlak, L. Waliś, *Modification of zirconium alloy surface using high intensity pulsed plasma beams,* Archives of Materials Science and Engineering, Feb 2016

[14] M. Gryziński: *Koncepcja prętowego (magnetoelektrycznego) działa plazmowego "DP".* Instytut Badań Jądrowych, Raport INR No **711**/XVIII/PP, Warszawa, maj, 1966

[15] M. Gryziński, *A new device for creating a strongly focused hot plasma jet – Rod Plasma Injector (RPI).* Nukleonika, vol. **XIV**, No. 7-8, (1969) 679-705

[16] J. Nowikowski, L. Jakubowski, *Investigations of RPI in dynamic gas conditions.* Nukleonika, vol. **XXI**, No. 11-12, (1976) 1227-1240

[17] K. Malinowski, „*Experimental Investigation and Computer Simulations of an Ion Emission of the RPI-IBIS Plasma Accelerator*", PhD Thesis, National Center for Nuclear Research, 2012.

[18] E. Skladnik-Sadowska, M. Sadowski, J. Baranowski, *Investigation of convergent deuteron beams within a penetrable electrode system.* Proc. 15th European Conf. on Controlled Fusion and Plasma Heating, Dubrovnik (1988), vol. **12B**, Part II, 633-636

[19] M.J. Sadowski, J. Baranowski, E. Skladnik-Sadowska, V.N. Borisko, O.V. Byrka, V.I. Tereshin, A.V. Tsarenko, *Characterization of pulsed plasma-ion streams emitted from RPI-type devices applied for material engineering,* Applied Surface Science **238** (2004) 433–437

[20] R. Kwiatkowski, "*Preliminary Measurements of Energy Distribution of Ions Emitted by IBIS-II Plasma Source*", NCNR Internal Report, National Center for Nuclear Research, 2018.

[21] Nuclear Data Services, *Stopping Power of Matter for Ions*, International Atomic Energy Agency, https://www-nds.iaea.org/stopping/stopping_timg.html

[22] A. Horodeński, *Evaluation of Pulse Shape of Ion Beams Produced by the Ionotron-Type Ion Sources*, phys. stat. sol. (a) **112**, 821 (1989)

[23] *New Semiconductor Materials. Characteristics and Properties,* Ioffe Physico-Technical Institute, http://www.ioffe.ru/SVA/NSM/Semicond/Si/thermal.html

[24] N. M. Ravindra, B. Sopori, O. H. Gokce, S. X. Cheng, A. Shenoy, L. Jin, S. Abedrabbo, W. Chen, Y. Zhang, *Emissivity Measurements and Modeling of Silicon-Related Materials: An Overview*, International Journal of Thermophysics, Vol. **22**, No. 5, September 2001

[25] *Vapor Pressure Calculator*, Technische Universität Wien, Institut für Angewandte Physik, https://www.iap.tuwien.ac.at/www/surface/vapor_pressure

[26] *A Simple Sputter Yield Calculator*, Technische Universität Wien, Institut für Angewandte Physik, https://www.iap.tuwien.ac.at/www/surface/sputteryield.